\documentclass[twocolumn,aps,showpacs,prb,tightenlines,amsmath,amssymb]{revtex4}
\usepackage{graphicx}
\usepackage{amssymb}
\usepackage{slashed}
\usepackage{dcolumn}
\usepackage{amsmath}
\usepackage{bm}
\usepackage{colordvi}
\usepackage{mathrsfs}
\makeatletter

\newcommand{\Rmnum}[1]{\expandafter\@slowromancap\romannumeral #1@}
\makeatother

\begin{document}

\title{Optical response of Higgs mode in superconductors at clean limit: formulation through Eilenberger equation and Ginzburg-Landau Lagrangian}   

\author{F. Yang}
\email{yfgq@mail.ustc.edu.cn.}

\affiliation{Hefei National Laboratory for Physical Sciences at
Microscale, Department of Physics, and CAS Key Laboratory of Strongly-Coupled
Quantum Matter Physics, University of Science and Technology of China, Hefei,
Anhui, 230026, China}

\author{M. W. Wu}
\email{mwwu@ustc.edu.cn.}

\affiliation{Hefei National Laboratory for Physical Sciences at
Microscale, Department of Physics, and CAS Key Laboratory of Strongly-Coupled
Quantum Matter Physics, University of Science and Technology of China, Hefei,
Anhui, 230026, China}

\date{\today}

\begin{abstract}

 Both macroscopic Ginzburg-Landau Lagrangian and microscopic gauge-invariant kinetic equation suggest a finite Higgs-mode generation in the second-order optical response of superconductors at clean limit, whereas the previous derivations through the path-integral approach and Eilenberger equation within the Matsubara formalism failed to give such generation. The crucial treatment leading to this controversy lies at an artificial scheme that whether the external optical frequency is taken as continuous variable or bosonic Matsubara frequency to handle the gap dynamics within the Matsubara formalism. To resolve this issue, we derive the effective action of the superconducting gap near $T_c$ in the presence of the vector potential through the path-integral approach, to fill the long missing blank of the microscopic derivation of the Ginzburg-Landau Lagrangian in superconductors. It is shown that only by taking optical frequency as continuous variable  within the Matsubara formalism, can one achieve the fundamental Ginzburg-Landau Lagrangian, and in particular, the finite Ginzburg-Landau kinetic term leads to a finite Higgs-mode generation at clean limit. To further eliminate the confusion of the Matsubara frequency through a separate framework, we apply the Eilenberger equation within the Keldysh formalism, which is totally irrelevant to the Matsubara space. By calculating the gap dynamics in the second-order response, it is analytically proved that the involved optical frequency is a continuous variable rather than bosonic Matsubara frequency, causing a finite Higgs-mode generation at clean limit.

\end{abstract}


\maketitle 

\section{Introduction}

In the past few decades, the Higgs mode in the field of superconductivity, which describes the amplitude fluctuation $\delta|\Delta|$ of the superconducting order parameter $\Delta$, has attracted much attention. This collective excitation corresponds to the radial excitation in the Mexican-hat potential of free energy\cite{Am0}, and hence, exhibits a gapful energy spectrum at a long wavelength\cite{OD1,OD2,OD3,pm5,Am0,Am5,Am6,Am12,AK2}. Owing to the advanced ultrafast terahertz technique in nonlinear optics, the Higgs mode has been experimentally observed and identified as the origin of the excited superfluid-density oscillation $\delta\rho_s$ in the second-order harmonic generation\cite{NL7,NL8,NL9,NL10,NL11,DHM2,DHM3}. The nonlinear optics in superconductors has since stimulated a lot of experimental interest and inspired a great deal of theoretical studies. 

Whereas the existing and growing experimental observations exhibit a very convincing evidence, the theoretical descriptions concerning the Higgs-mode excitation in the literature are filled with controversies, which are detrimental to the understanding of the related experimental findings. The central issue lies at a question that with the conventional light-matter interactions $H_d={{\hat {\bf p}}\cdot{e{\bf A}}}\tau_0/{m}$ (current-vertex-related drive effect) and $H_p=e^2A^2\tau_3/(2m)$ (density-vertex-related pump effect) by the vector potential ${\bf A}$\cite{Ba0,G1,GIKE2}, whether the Higgs mode can be optically excited at clean limit. Here, $\tau_{i=0,1,2,3}$ denote the Pauli matrices in Nambu space. 

Concerning this issue, although the early stage of works through the Bloch\cite{Am1,Am2,Am7,Am9,Am11,Am14,Am15,NL7,NL8,NL9,NL10,NL11} or Liouville\cite{Am3,Am4,Am8,Am10,Am16} equation within the Anderson pseudospin picture\cite{As} revealed an excited fluctuation of the order parameter by pump effect $H_p$, a later symmetry analysis\cite{symmetry} implies that the pumped order-parameter fluctuation by density-vertex-related $H_p$ is a phase fluctuation rather than the claimed amplitude one\cite{GIKE2}, as the revealed correlation between amplitude (phase) mode and $H_p$ in Ref.~\onlinecite{symmetry} is zero (nonzero). The path-integral approach is naturally capable of distinguishing the excitations of phase and Higgs modes by deriving the corresponding effective actions, and includes both pump and drive effects. Using this approach, Cea {\em et al.} derived a vanishing Higgs-mode generation in second-order response at clean limit\cite{Cea1,Cea2,Cea3}, in inconsistency with the previous experimental understanding\cite{NL7,NL8,NL9}. To explain the experimental findings, Cea {\em et al.} pointed out\cite{Cea1,Cea2,Cea3} that the density-vertex-related pump effect $H_p$ can excite a finite fluctuation $\delta{n}$ of charge density $n$, so they speculated that the experimentally observed superfluid-density oscillation $\delta\rho_s$ is attributed to charge-density fluctuation $\delta{n}$ rather than the Higgs mode $\delta|\Delta|$, since the superfluid density $\rho_s\propto{n}|\Delta|^2$.

In the several polarization-resolved measurements afterwards\cite{NL10,NL11,DHM2,DHM3}, an isotropic second-order harmonic signal was timely reported and provides a firm evidence to rule out the possible charge-density fluctuation, as the theoretically predicted response of the Higgs mode (charge-density fluctuation) is isotropic (anisotropic)\cite{Cea1}. Since then, it is believed that the Higgs-mode generation is zero at clean limit and one has to reply on impurity scattering to mediate the Higgs-mode generation\cite{FHM,Am16,ImR1,ImR2,ImR3,Silaev}. In this situation, to take account of the microscopic scattering, Silaev applied the Eilenberger equation\cite{Eilen} within the Matsubara formalism\cite{Silaev,Silaev0}, which solely includes the current-vertex-related drive effect $H_d$. He also derived a vanishing Higgs-mode generation at clean limit, but with impurities, a finite one to dominate over the charge-density fluctuation is obtained. Particularly, Silaev showed that the impurity scattering only mediate the Higgs-mode excitation and is incapable of causing the damping of this collective excitation\cite{Silaev}, so the increase of the impurity density can enhance the optical signal of Higgs mode.

Meanwhile, using a gauge-invariant kinetic equation approach\cite{GIKE1,GIKE2,GIKE3} with complete electromagnetic effect and microscopic scattering, Yang and Wu derived totally opposite results. They obtain a finite Higgs-mode generation contributed by the drive effect at clean limit\cite{GIKE2}, and show that the charge-density fluctuation in fact vanishes in the second-order response, as a consequence of the charge conservation and forbidden second-order harmonic current in systems with the spacial inversion symmetry. The revealed Higgs-mode generation at clean limit can capture the experimental observation well\cite{GIKE2}, and in particular, a finite damping/lifetime of the Higgs-mode excitation by impurities is also derived\cite{GIKE3}, providing a possible origin for the experimentally observed broadening of the Higgs-mode resonance signal as well as the fast Higgs-mode damping after optical excitation. This damping agrees with the analysis of Heisenberg equation of motion since the Higgs-mode excitation and electron-impurity interaction are non-commutative in Nambu space.

This finite Higgs-mode generation has therefore been in sharp contrast to the aforementioned vanishing one from Eilenberger equation and path-integral approach. Actually, at clean limit, the finite Higgs-mode generation in second-order response of superconductors is a direct consequence of the Ginzburg-Landau Lagrangian\cite{EPM}. This is because that from the time-dependent Ginzburg-Landau superconducting Lagrangian at clean limit, which was proposed by Pekker and Varma through the symmetry analysis and Lorentz invariance from the Landau phase-transition theory\cite{Am6}, one can directly reveal the equation of motion of the Higgs mode by considering the amplitude fluctuation of the Landau order parameter. Then, a finite Higgs-mode generation in the second-order response is immediately obtained\cite{EPM}. This directly leads to a particularly bizarre question that why both path-integral approach\cite{PI2GL} and Eilenberger equation\cite{Ba20} can recover the Ginzburg-Landau equation but reach a zero Higgs-mode generation.

To resolve this controversy, by re-examining the previous derivations within the path-integral approach in Refs.~\onlinecite{Cea1,Cea2,Cea3} and Eilenberger equation in Ref.~\onlinecite{Silaev}, it is pointed out by Yang and Wu\cite{EPM} that both previous derivations contain flaws. Specifically, the previous works\cite{Cea1,Cea2,Cea3} within the path-integral approach only kept the perturbation expansion of the action up to second order, whereas the essential coupling of the Higgs mode to the second order of the drive effect $H_d$ emerges in the third-order perturbation expansion of the action\cite{EPM}. So this coupling and hence finite second-order harmonic generation of Higgs mode by the drive effect are excessively overlooked in Refs.~\onlinecite{Cea1,Cea2,Cea3}.  Within the Matsubara formalism, picking up this coupling in the path-integral approach, one can find the exactly same amplitude-response coefficient as the one derived from Eilenberger equation\cite{Silaev}:
\begin{eqnarray}\label{le1}
      \lambda_E\!\!&=&\!\!\frac{T}{2\Omega^2}\sum_{p_n}\Big[\frac{2}{\sqrt{(p_n\!-\!i\Omega)^2\!+\!\Delta_0^2}}\!-\!\frac{1}{\sqrt{(p_n\!-\!2i\Omega)^2\!+\!\Delta_0^2}}\nonumber\\
        &&\mbox{}-\!\frac{1}{\sqrt{p_n^2\!+\!\Delta_0^2}}\Big], 
\end{eqnarray}
where $p_n=(2n+1)\pi{T}$ represents the fermionic Matsubara frequencies.

Nevertheless, in Ref.~\onlinecite{Silaev}, the involved optical frequency $\Omega$ is taken as bosonic Matsubara frequency $i\Omega_m$, leading to a vanishing response coefficient $\lambda_E$ in Eq.~(\ref{le1}) strongly against the finite one from gauge-invariant kinetic equation\cite{EPM,GIKE2} and Ginzburg-Landau Lagrangian\cite{EPM}. Moreover, because of this treatment, the prefactor $1/\Omega^2$ in Eq.~(\ref{le1}) causes an undefined singularity at zero frequency, and an unphysical discontinuity between cases at $\Omega=0$ and $\Omega\rightarrow0$ emerges\cite{EPM}. In contrast, the previous work in Ref.~\onlinecite{EPM} takes $\Omega$ as continuous variable. A finite Higgs-mode generation at clean limit in agreement with the Ginzburg-Landau Lagrangian is then derived, and the obtained $\lambda_E$ from both Eilenberger equation and path-integral approach becomes exactly same as the one from gauge-invariant kinetic equation.

Actually, the Matsubara formalism is developed as an auxiliary-function technique in the finite-temperature Green function approach. In this framework, whether taking the external optical frequency $\Omega$ as continuous variable or bosonic Matsubara frequency $i\Omega_m$ can not be self-justified by method itself. The treatment of $\Omega$ as $i\Omega_m$ in the calculations of conductivity and dielectric function in normal metals\cite{mt} can be cross-justified by many other methods irrelevant to Matsubara space, such as equation of motion, zero-temperature Green function, Boltzmann equation as well as Keldysh Green function approaches, whereas such justification in the gap dynamics of superconductors has not been performed in the literature so far.  Physically, any treatment leading to result strongly against the Ginzburg-Landau superconducting Lagrangian can not be correct. Nevertheless, this justification in superconductors has been challenged, arguing that the Ginzburg-Landau superconducting Lagrangian is a phenomenological model near $T_c$ and is unimportant in microscopic studies, as the microscopic derivation of this Lagrangian is absent in the literature. Theoretically, this Lagrangian is a fundamental model by symmetry analysis and Lorentz invariance as well as Landau phase-transition theory, whereas to fill the long missing blank of the microscopic derivation, in the present work, we derive the Lagrangian in superconductors through the basic path-integral approach. In addition to this physical justification, a natural and rigorous framework developed in the literature to eliminate the confusion of the Matsubara frequency in superconductors is to perform the formulation within the Keldysh formalism\cite{QA1}, which is a well-established systematic approach for studying non-equilibrium properties and is totally irrelevant to Matsubara space. We therefore apply the Eilenberger equation within the Keldysh formalism to calculate the gap dynamics at clean limit to cross-justify the treatment of the external optical frequency.

Specifically, through the path-integral approach within the Matsubara formalism, we derive the effective action of superconducting gap near $T_c$ in the presence of the vector potential. We show that to recover the Ginzburg-Landau kinetic term, one needs to keep the perturbation expansion of the action up to the fourth order and formulate the fourth-order correlation coefficient.  Particularly, during our calculation of the correlation coefficient, it is shown that only by taking optical frequency as continuous variable, one can recover the finite coefficient in the Ginzburg-Landau kinetic term. Then, if one considers the gap fluctuation to obtain the equation of motion of the Higgs mode from the Ginzburg-Landau Lagrangian, it is clearly seen that the finite coefficient in the Ginzburg-Landau kinetic term directly leads to the finite response coefficient in the equation of motion of the Higgs mode. 

Furthermore, with the vector potential alone, we apply the Eilenberger equation within the Keldysh formalism to derive the equation of motion of the Higgs mode at clean limit. It is established in the literature that the Keldysh Green function can be written as a function of the retarded and advanced ones through a general relation via the distribution function. We prove that this relationship makes the Keldysh Green function directly satisfying the normalization condition of the Eilenberger equation. Then, we solve the retarded Green function and distribution function to obtain the Keldysh Green function. With the derived Keldysh Green function, we obtain the equation of motion of the Higgs mode, which is exactly same as the one derived through Eilenberger equation within the Matsubara formalism\cite{Silaev}. In contrast, as our derivation is performed in the Keldysh formalism and is irrelevant to the Matsubara space, it is clearly shown that the involved optical frequency in the response coefficient in the equation of motion of the Higgs mode is a continuous variable rather than the Matsubara frequency. Particularly, with the continuous optical frequency, the response coefficient in the equation of motion of the Higgs mode does not vanish, leading to a finite Higgs mode generation at clean limit.

\section{Hamiltonian}
\label{sec-H}

We first present the general Bogoliubov-de Gennes Hamiltonian of the conventional $s$-wave superconductors in the presence of the electromagnetic potential\cite{G1,Ba0,AK,GIKE2}: 
\begin{equation}
  H\!=\!{\int}{d{\bf x}}~\psi^{\dagger}(x)
  \left(\begin{array}{cc}
   \xi_{{\hat{\bf p}}-e{\bf
        A}}+e\phi  & \Delta(x) \\
   \Delta^*(x) & -\xi_{{\hat{\bf p}}+e{\bf
        A}}-e\phi
  \end{array}\right)\psi(x). \label{BdG}
\end{equation}
Here, $\psi(x)=[\psi_{\uparrow}(x),\psi^{\dagger}_{\downarrow}(x)]^T$ is the field operator in the Nambu space and $x=(t,{\bf x})$ represents the space-time vector; $\phi$ and ${\bf A}$ denote the scalar and vector potentials, respectively; the momentum operator ${\hat {\bf p}}=-i\hbar{\bm \nabla}$ and $\xi_{\hat{\bf p}}={{\bf{\hat p}}^2}/({2m})-\mu$ with $m$ being the effective mass and $\mu$ denoting the chemical potential. It is noted that the scalar potential can be generally written as $\phi={\bar \phi}_0+{\bf E}_{\phi}\cdot{\bf x}+\delta\phi$, where ${\bar \phi}_0$ denotes the effect of the electric voltage; ${\bf E}_{\phi}\cdot{\bf x}$ concerns the drive effect by electric field; $\delta\phi$ is the induced scalar potential related to the long-range Coulomb interaction (i.e., Hartree field caused by charge density fluctuation)\cite{Ba0,AK,GIKE2}. 

In consideration of the phase $\delta\theta(x)$ and amplitude $\delta|\Delta(x)|$ fluctuations around the equilibrium gap $\Delta_0$, the superconducting order parameter reads:
\begin{equation}
  \Delta(x)=|\Delta(x)|e^{i\delta\theta(x)}=[\Delta_0+\delta|\Delta(x)|]e^{i\delta\theta(x)}.\end{equation} 

It is established that the phase mode $\delta\theta$ in Hamiltonian above can be effectively removed by a unitary transformation\cite{gi0,gi1,AK}
\begin{equation}\label{ut}
  \psi(x){\rightarrow}e^{i\tau_3\delta\theta(x)/2}\psi(x),
\end{equation}
and then, one has
\begin{equation}
  H=\!\!\!{\int}{d{\bf x}}~\psi^{\dagger}(x)(H_0+H_{\rm LM})\psi(x),
\end{equation}
where the free BCS Hamiltonian $H_0$ is written as 
\begin{equation}
H_0=\xi_{\bf {\hat p}}\tau_3+|\Delta(x)|\tau_1,  
\end{equation}  
and the light-matter interaction reads
\begin{equation}
  H_{\rm LM}=\frac{{\bf p}_s\cdot{\hat{\bf p}}}{m}+\frac{p_s^2}{2m}\tau_3+\mu_{\rm eff}\tau_3
\end{equation}
with the gauge-invariant superconducting momentum {\small ${\bf p}_s={\bm \nabla}\delta\theta/2-e{\bf A}$} and effective field {\small $\mu_{\rm eff}=e\phi+\partial_{t}\delta\theta/2$}. 

It has been revealed in the literature\cite{AK,GIKE2} that thanks to the Coulomb screening, in the linear regime, at long-wavelength limit, the induced scalar potential $e\delta\phi$ cancels the original longitudinal part $e{\bar \phi}_0+\partial_{t}\delta\theta/2$ in the effective field {\small $\mu_{\rm eff}=e{\bar \phi}_0+e{\bf E}_{\phi}\cdot{\bf x}+e\delta\phi+\partial_{t}\delta\theta/2$}, leaving only the drive effect of scalar-potential-induced electric field $E_{\phi}$. In general, the inclusion of this retained effect from scalar potential is essential to capture the optical-electric-field effect in a gauge-invariant manner. Whereas considering the fact that the vector potential characterizes the Meissner effect/Ginzburg-Landau kinetic term in addition to the optical-electric-field effect\cite{GIKE1,GIKE2,EPM,PYW}, in the present work, we only focus on the electromagnetic effect from vector potential, similar to the previous works by Cea {\em et al.}\cite{Cea1,Cea2,Cea3} and Silaev\cite{Silaev}. The light-matter interaction then becomes
\begin{equation}\label{LMF}
H_{\rm LM}=\frac{{\bf p}_s\cdot{\hat{\bf p}}}{m}+\frac{p_s^2}{2m}\tau_3.
\end{equation}  
Moreover, it has been established in the literature\cite{Am0,Ba0,AK,AK2,pm0,pm5,pi1,pi4,GIKE2} that the linear response of the phase mode, as a scalar quantity, responds to the longitudinal electromagnetic field and hence experiences the Coulomb screening. The resonance pole of this response is then effectively lifted from the original gapless spectrum up to the high-energy plasma frequency $\omega_p$ as a consequence of the Anderson-Higgs mechanism\cite{AHM}, and hence, no effective linear response of the phase mode occurs at frequency $\Omega\ll\omega_p$. Because of this effect, the linear response of the phase mode cancels the unphysical longitudinal vector potential in ${\bf p}_s$, and the superconducting momentum that appears in the previous theoretical descriptions in the literature only involves the physical transverse vector potential. The light-matter interaction in Eq.~(\ref{LMF}) then consists of the drive effect $H_d$ and pump one $H_p$

\section{Derivation of Ginzburg-Landau Lagrangian}

In this section, we present the derivations of the time-dependent Ginzburg-Landau Lagrangian and its non-equilibrium variation (equation of motion of the Higgs mode). Specifically, 
the action of superconductors in the presence of vector potential after the Hubbard-Stratonovich transformation is written as\cite{Ba0,pi1,pi4,PI2GL} 
\begin{eqnarray}
&&\!\!\!\!\!S[\psi,\psi^*]=\!\!\int{dx}\bigg[\sum_{s=\uparrow,\downarrow}\!\!\psi^*_s(x)(i\partial_t\!-\!\xi_{\hat {\bf p}-e{\bf A}})\psi_s(x)\nonumber\\
&&\mbox{}+\!\psi^*_{\uparrow}(x)\psi^*_{\downarrow}(x)\Delta(x)\!+\!\psi_{\downarrow}(x)\psi_{\uparrow}(x)\Delta^*(x)\!-\!\frac{|\Delta(x)|^2}{U}\bigg].~~~~~~  
\end{eqnarray}
Considering the spatial dependence of the gap,  one has $|\Delta(x)|=\sum_{\bf q}|\Delta_{\bf q}|e^{i{\bf q}\cdot{\bf x}}$ in Fourier space. Then, applying the unitary transformation in Eq.~(\ref{ut}), the above action in momentum space becomes
\begin{eqnarray}
&&\!\!\!\!S[\psi,\psi^*]=\int{dt}\Big\{\sum_{\bf kq}\big[\psi^*_{{\bf k+\frac{q}{2}}\uparrow}(i\partial_t\!-\!\xi_{\bf k+\frac{q}{2}+p_s})\psi_{{\bf k+\frac{q}{2}}\uparrow}\nonumber\\
    &&\mbox{}+\psi^*_{{\bf -k+\frac{q}{2}}\downarrow}(i\partial_t\!-\!\xi_{\bf -k+\frac{q}{2}+p_s})\psi_{{\bf -k+\frac{q}{2}}\downarrow}+(\psi_{{\bf -k+\frac{q}{2}}\downarrow}\psi_{{\bf k+\frac{q}{2}}\uparrow}\nonumber\\
  &&\mbox{}+\psi_{{\bf k+\frac{q}{2}}\uparrow}^*\psi_{{\bf -k+\frac{q}{2}}\downarrow}^*)|\Delta_{\bf q}|\big]-\sum_{\bf q}\frac{|\Delta_{\bf q}|^2}{U}\Big\},
\end{eqnarray}
and in Nambu space, one finds the action related to gap:
\begin{eqnarray}
S[\psi,\psi^*]=\!\!\int{dt}\sum_{\bf q}\!\Big[\sum_{\bf k}\psi^{\dagger}_{\bf kq}(G_{0}^{-1}\!-\!\Sigma)\psi_{\bf kq}\!-\!\frac{|\Delta_{\bf q}|^2}{U}\Big].
\end{eqnarray}
Here, the field operator $\psi_{\bf kq}^{\dagger}=(\psi_{{\bf k+\frac{q}{2}}\uparrow}^*,\psi_{{\bf -k+\frac{q}{2}}\downarrow})$; $G_{0}^{-1}=i\partial_t-\xi_{\bf k}\tau_3$ which gives the Green function $G_0(p)=(p_0-\xi_{\bf k}\tau_3)^{-1}$ in frequency space with the four-vector momentum $p=(p_0,{\bf k})$; the self-energy reads
\begin{equation}
  \Sigma=|\Delta_{\bf q}|\tau_1+\frac{{\bf k}\cdot({\bf q/2+p_s})}{m}+\frac{({\bf q/2+p_s})^2}{2m}\tau_3.
\end{equation}  

Considering the small gap near critical temperature $T_c$ as well as the weak strength and spatial variation of the vector potential, the self-energy can be treated as small quantity. Then, after the standard integration over the Fermi field, one has
\begin{equation}
S=S_0\!-\sum_{n=1}^{\infty}\frac{1}{n}{\rm {\bar Tr}}[(G_0\Sigma)^n]\!-\!\int{dt}\sum_{\bf q}\frac{|\Delta_{\bf q}|^2}{U}.
\end{equation}

To derive the Lagrangian related to the superconductivity, one needs to formulate
the expansions of the action with respect to the fourth order of the self-energy (i.e., keep the expansions up to $n=4$). Then, with expansions up to $n=4$, by only keeping the terms related to the gap, one can obtain the effective action $S_s=\int{dq}\mathscr{L}$ with the frequency-momentum four vector $q=(\Omega,{\bf q})$ and the Lagrangian of the superconductivity:
\begin{eqnarray}
  \mathscr{L}\!\!&=&\!\!-\chi_1|\Delta_{\bf q}|\!-\!\big(\frac{1}{2}\chi_{11}+\frac{1}{U}\big)|\Delta_{\bf q}|^2\!-\!\chi_{13}\frac{({\bf q/2\!+\!p_s})^2|\Delta_{\bf q}|}{2m}\nonumber\\
  &&\!\!\mbox{}-\!\chi_{111}\frac{|\Delta_{\bf q}|^3}{3}\!-\!\chi_{100}\frac{k_F^2({\bf q/2\!+\!p_s})^2|\Delta_{\bf q}|}{3m^2}\!-\!\frac{\chi_{1111}|\Delta_{\bf q}|^4}{4}\nonumber\\
  &&\!\!\mbox{}-\!\chi_{113}\frac{({\bf q/2\!+\!p_s})^2|\Delta_{\bf q}|^2}{2m}\!-\!\chi_{1113}\frac{({\bf q/2\!+\!p_s})^2|\Delta_{\bf q}|^3}{2m}\nonumber\\
  &&\!\!\mbox{}-\!(\chi_{1100}+\chi_{0110}+\chi_{1010})\frac{k_F^2({\bf q/2\!+\!p_s})^2|\Delta_{\bf q}|^2}{6m^2}. \label{LG1}
\end{eqnarray}
Here, we have neglected the terms proportional to odd orders of ${{\bf k}\cdot({\bf q/2+p_s})}$ as these anisotropic terms vanish after the summation of the momentum ${\bf k}$. The correlation coefficients are determined by 
\begin{eqnarray}
\chi_{i}\!\!&=&\!\!\!\sum_p{\rm Tr}[G_0(p)\tau_j], \label{chii}  \\
\chi_{ij}\!\!&=&\!\!\!\sum_p{\rm Tr}[G_0(p\!+\!q)\tau_iG_0(p)\tau_j], \label{chiij} \\
  \chi_{ijk}\!\!&=&\!\!\!\sum_p{\rm Tr}[G_0(p\!+\!2q)\tau_iG_0(p\!+\!q)\tau_jG_0(p)\tau_k],\label{chiijk}\\
 \chi_{ijkl}\!\!&=&\!\!\!\sum_p{\rm Tr}[G_0(p\!+\!q)\tau_iG_0(p)\tau_jG_0(p\!+\!q)\tau_kG(p)\tau_l].\label{chiijkl}~~~~~
\end{eqnarray}
Since the Green function only consists of the $\tau_0$ and $\tau_3$ components, one immediately finds $\chi_1=\chi_{13}=\chi_{111}=\chi_{100}=\chi_{1113}=0$. Moreover, the coefficient $\chi_{113}$ vanishes due to the particle-hole symmetry, and the forth-order correlation coefficient $\chi_{1010}=\chi_{0110}+\chi_{1100}$  (refer to Appendix). Then, one finds an embryonic form of the Ginzburg-Landau Lagrangian of superconductors: 
\begin{equation}
  \mathscr{L}=-\chi_p|\Delta_{\bf q}|^2-\frac{\beta_p|\Delta_{\bf q}|^4}{2}-\frac{\lambda_p({\bf q/2\!+\!p_s})^2|\Delta_{\bf q}|^2}{m}. \label{LG2}
\end{equation}
with the parameters $\chi_p={\chi_{11}}/{2}+{1}/{U}$ and $\beta_p=\chi_{1111}/2$ as well as $\lambda_p=k_F^2\chi_{1010}/(3m)$ .

Within the Matsubara formalism $[p=(ip_n,{\bf k})]$, the retained correlation coefficients in Eq.~(\ref{LG2}) are given by (refer to Appendix~\ref{ADCC})
\begin{eqnarray}
   \chi_p&=&\alpha_p-\Omega^2\gamma_p/2,\\
  \beta_p&=&-\!\!\!\!\sum_{ip_{n>0},\eta=\pm}\frac{2\pi{iD}}{\beta}\frac{2}{(2ip_n\!+\!\eta\Omega)^3},\label{betapp}\\
  \lambda_p&=&\!\!\!\!\!\!\!\sum_{ip_{n>0},\eta=\pm}\!\!\frac{k_F^2}{3m}\frac{2\pi{i}D}{\beta\Omega^2}\Big[\frac{4}{2ip_n\!+\!\eta\Omega}\!-\!\frac{1}{ip_n}\!-\!\frac{1}{ip_n\!+\!\eta\Omega}\Big],~~~~~\label{f1010}
\end{eqnarray}
with parameters: 
\begin{eqnarray}
  \alpha_p&=&\!\!D\int^{\omega_D}_{-\omega_D}d\xi_{\bf k}\frac{\tanh(\frac{\beta_c\xi_{\bf k}}{2})\!-\!\tanh(\frac{\beta\xi_{\bf k}}{2})}{2\xi_{\bf k}}\!=\!D\ln\frac{T}{T_c},~~~~\label{apf}\\
  \gamma_p&=&\!\!\!\!\!\!\!\sum_{ip_{n>0},\eta=\pm}\!\!\frac{\pi{i}D}{\beta\Omega^2}\Big[\frac{4}{2ip_n\!+\!\eta\Omega}\!-\!\frac{1}{ip_n}\!-\!\frac{1}{ip_n\!+\!\eta\Omega}\Big].\label{fgp}
\end{eqnarray}
Here, $D$ and $\omega_D$ denote the density of states and Debye frequency, respectively; $f(x)$ stands for the Fermi distribution; $\beta=k_BT$ and $\beta_c=k_BT_c$ with $k_B$ being the Boltzmann constant.

At low frequency ($\Omega<{k_BT_c}$), one finds the specific parameters:
\begin{eqnarray}
  \gamma_p&=&\beta_p\approx\sum_{n>0}\frac{\pi{D}}{\beta{p_n^3}}=\frac{7D\zeta(3)}{8({\pi}T)^2},~~~~\label{bpf}\\
  \lambda_p&\approx&\sum_{n>0}\frac{k_F^2}{3m}\frac{2\pi{D}}{4\beta}\frac{4}{p_n^3}=\frac{k_F^2}{3m}\frac{7D\zeta(3)}{(2{\pi}T)^2},\label{lpf}
\end{eqnarray}
with $\zeta(x)$ being the Riemann zeta function. Then, through the transformation into time-space coordinate, the time-dependent Ginzburg-Landau Lagrangian of superconductors is obtained: 
\begin{equation}
\mathscr{L}\!=\!\frac{\gamma_p|i\partial_t\Delta|^2}{2}\!-\!\Big(\alpha_p|\Delta|^2\!+\!\frac{\beta_p|\Delta|^4}{2}\!+\!\frac{\lambda_p|({\bm \nabla}\!-\!2ie{\bf A})\Delta|^2}{4m}\Big),  \label{GLLE}
\end{equation}
which is exactly same as the one obtained by Pekker and Varma through symmetry analysis and Lorentz invariance from the general Ginzburg-Landau free energy\cite{Am6} as it should be. Moreover, the parameters $\alpha_p$ [Eq.~(\ref{apf})] and $\beta_p$ [Eq.~(\ref{bpf})] as well as $\lambda_p$ [Eq.~(\ref{lpf})] here are exactly same as the obtained Landau parameters in the previous work by Gorkov\cite{G1} in which the Ginzburg-Landau equation is derived from Gorkov equation.

It is noted that in the derivation of the gap dynamics within the Matsubara formalism in the present work, we take the optical frequency $\Omega$ as continuous variable rather than bosonic Matsubara frequency $i\Omega_m$, and only with continuous $\Omega$ in this circumstance, can one recover the Ginzburg-Landau Lagrangian as demonstrated above. Actually, as seen from Eq.~(\ref{f1010}), if $\Omega$ is taken as $i\Omega_m=2mi\pi{T}$, different irrational response coefficients are obtained at cases with odd and even $m$.  For odd $m$, a divergent pole ($2n+1=m$ and $\eta=-1$) which is unable to circumvent emerges in the formulation of the first term on the right-hand side of Eq.~(\ref{f1010}). As for even $m$, through the frequency displacement in the Matsubara frequency summation, the response coefficient $\lambda_p$ directly vanishes for nonzero $m$, but due to the prefactor $1/\Omega^2$ in Eq.~(\ref{f1010}), there exists an undefined singularity at zero frequency. Then, an unphysical discontinuity between $\Omega=0$ and $\Omega\rightarrow0$ emerges. The difference between cases with odd and even $m$ in bosonic Matsubara frequency $i\Omega_m=2mi\pi{T}$ is totally unreasonable within the Matsubara formalism, and neither of them can recover the finite Landau parameter at low frequency. Clearly, any treatment that leads to consequence strongly against the fundamental model of symmetry analysis and Lorentz invariance as well as Landau phase-transition theory can not be correct. All of the irrationalities here simply suggest that the treatment of taking $\Omega$ as $i\Omega_m$ can not be correct in the derivation of gap dynamics in superconductors.

Furthermore, from the Lagrangian above, with the equilibrium gap $\Delta_0=\sqrt{-\alpha_p/\beta_p}$, by considering the gap fluctuation $\delta|\Delta|$ through $|\Delta|=\Delta_0+\delta|\Delta|$, its equation of motion at long-wavelength limit is directly obtained:
\begin{equation}\label{GL-HME}
\big[(2\Delta_0)^2-\partial_t^2\big]\delta|\Delta|=-\frac{\lambda_p}{\gamma_p}\frac{e^2A^2}{m}2\Delta_0.  
\end{equation}
Then, one immediately finds the Higgs-mode energy spectrum (i.e., resonance pole) $\omega_H=2\Delta_0$ on the left-hand side of the equation above, and in particular, a finite second-order response of Higgs mode at clean limit on the right-hand side of the equation above. It is pointed out that although the derivation of the equation of motion of the Higgs mode here is based on the small gap and only holds near $T_c$, the serious derivation in regime extending to $T=0$ through three different microscopic approaches including the gauge-invariant kinetic equation, Eilenberger equation as well as path-integral approach also obtains the similar equation of motion as a consequence of the renormalization group (scaling) theory or basic local Abelian $U(1)$ model (complex scalar field coupled to an electromagnetic potential) in the field theory.

\section{Derivation of Higgs-mode response through Eilenberger equation}

To eliminate the confusion of the Matsubara frequency from a separate framework in addition to the physical justification above, in this section, we perform the formulation of the gap dynamics within the Keldysh formalism, which is totally irrelevant to Matsubara space. Specifically, we apply the Eilenberger equation within the Keldysh formalism to derive the second-order optical response of the Higgs mode at clean limit. The Eilenberger equation\cite{Eilen,Eilen1} is derived from the basic Gorkov equation of $\tau_3$-Green function through the quasiclassical approximation\cite{QA1}:
\begin{equation}\label{QG1}
g^{R/K/A}_{{\bf R,k_F}}(t,t')=\frac{i}{\pi}\!\!\int{d\xi_{\bf k}}\!\!\int{d{\bf r}}\tau_3G^{R/K/A}(x,x')e^{-i{\bf k}\cdot({\bf x}-{\bf x'})}.  
\end{equation}
Here, ${\bf R}=({\bf x+x'})/{2}$ represents the center-of-mass spatial coordinate; the retarded (R), advanced (A) and Keldysh (K) Green functions are defined by\cite{QA1,Eilen1}
\begin{eqnarray}
  G^R(x,x')&=&-i\langle\{\psi(x),\psi^{\dagger}(x')\}\rangle\theta(t-t'),\\
  G^A(x,x')&=&i\langle\{\psi(x),\psi^{\dagger}(x')\}\rangle\theta(t'-t),\\
  G^K(x,x')&=&-i\langle[\psi(x),\psi^{\dagger}(x')].
\end{eqnarray}

The Eilenberger equation within the Keldysh formalism at clean limit reads\cite{Eilen,Eilen1}:
\begin{equation}
  i\{\tau_3\partial_{t},{\hat g}\}_{t}-[(\Delta_0+\delta|\Delta|)\tau_1\tau_3,{\hat g}]_{t}+[e{\bf A}\cdot{\bf v}_F\tau_3,{\hat g}]_{t}\!=\!0, \label{ELE}
\end{equation}  
where the green function matrices ${\hat g}$ is defined as
\begin{equation}
  {\hat g}=\left(\begin{array}{cc}
    g^R & g^K \\
    0 & g^A
  \end{array}\right).  
\end{equation}
Here, the operators {\small $[X,{\hat g}]_{t}=X(t_1){\hat g}(t_1,t_2)-{\hat g}(t_1,t_2)X(t_2)$} and {\small $\{X,{\hat g}\}_{t}=X(t_1){\hat g}(t_1,t_2)+{\hat g}(t_1,t_2)X(t_2)$}.

Moreover, to guarantee the unique solution, the Eilenberger equation must be supplemented by the normalization condition\cite{Eilen,Eilen1}:
\begin{equation}
{\hat g}\circ{\hat g}=1,\label{nc}
\end{equation}
where the operator $\circ$ is defined by relation $A\circ{B}=\int{dt}A(t_1,t)B(t,t_2)$.

The corresponding gap equation is written as
\begin{equation}\label{ELGE}
\Delta_0+\delta|\Delta|=-iU{\rm Tr}[\langle{g^K_{\bf R,k_F}(t,t)}\rangle_F\tau_2/2],  
\end{equation}
with $\langle...\rangle_F$ denoting the angular average over the Fermi surface.

Considering an external optical field with ${\bf A}(t)={\bf A}_0e^{-i\Omega{t}}$, by self-consistently solving Eqs.~(\ref{ELE})-(\ref{ELGE}), one can formulate the Higgs-mode generation at clean limit. Specifically, in this circumstance, one can expand the quasiclassical Green function matrices as
\begin{equation}\label{RT}
{\hat g}={\hat g}^{(0)}+\delta{\hat g}^{(1)}+\delta{\hat g}^{(2)},
\end{equation}
with the $m$-th order response $\delta{\hat g}^{(m)}$ on the initial state ${\hat g}^{(0)}$. Correspondingly, the Higgs-mode generation $\delta|\Delta|=\delta|\Delta|^{(1)}e^{-i\Omega{t}}+\delta|\Delta|^{(2)}e^{-2i\Omega{t}}$ from Eq.~(\ref{ELGE}) with $\delta|\Delta|^{(1)}$ and $\delta|\Delta|^{(2)}$ being the excitations in first- and second-order optical responses, respectively. Particularly, the first-order response of Keldysh Green function must be anisotropic in momentum space, leading to a vanishing $\delta|\Delta|^{(1)}$ after the angular average over the Fermi surface in the gap equation. We then directly take $\delta|\Delta|^{(1)}=0$ for convenience.  

Consequently, the Eilenberger equation in Eq.~(\ref{ELE}) becomes a chain of equations:
\begin{equation}
\{\tau_3\partial_{t},\delta{\hat g}^{(1)}\}_{t}\!+\![\Delta_0\tau_2,\delta{\hat g}^{(1)}]_{t}\!-\!i[e{\bf A}\cdot{\bf v}_F\tau_3,{\hat g}^{(0)}]_{t}\!=\!0, \label{ELE1}
\end{equation}
\begin{eqnarray}
  &&\{\tau_3\partial_{t},\delta{\hat g}^{(2)}\}_{t}+[\Delta_0\tau_2,\delta{\hat g}^{(2)}]_{t}-i[e{\bf A}\cdot{\bf v}_F\tau_3,\delta{\hat g}^{(1)}]_{t}\nonumber\\
  &&\mbox{}+[\delta|\Delta|^{(2)}e^{-2i\Omega{t}}\tau_2,{\hat g}^{(0)}]_{t}\!=\!0, \label{ELE2}
\end{eqnarray}
and one can solve $\delta{\hat g}^{(1)}$ and $\delta{\hat g}^{(2)}$ in sequence with the given initial state ${\hat g}^{(0)}$. Then, with the obtained solution of the Keldysh Green function, one can further derive the response of the Higgs mode from the gap equation. 

\subsection{Solution of retarded Green function}
\label{SecGR}
           
In this part, we first solve the retarded Green functions from 
Eq.~(\ref{ELE}). From the expansion of Green function matrices in Eq.~(\ref{RT}), the retarded Green function is written as
\begin{eqnarray}
  g^R(t,t')&=&g^{R(0)}(t,t')+{\delta}g^{R(1)}(t,t')+{\delta}g^{R(2)}(t,t')\nonumber\\
  &=&\!\!\!\int\!\frac{dE}{2\pi}e^{iE(t'-t)}[g^{R(0)}(E)\!+\!{\delta}g^{R(1)}(E)e^{-i{\Omega}t}\nonumber\\
  &&\mbox{}+{\delta}g^{R(2)}(E){e^{-2i{\Omega}t}}].~~~~~
\end{eqnarray}  
The initial state of the retarded Green function has been established in the literature from Gorkov equation\cite{Eilen1,G1,Silaev}, and is written as
\begin{equation}\label{gr0}
g^{R(0)}(E)=\int\frac{d\xi_{\bf k}}{\pi}\frac{i\tau_3(E\!+\!\xi_{\bf k}\tau_3\!+\!\Delta_0\tau_1)}{(E\!+\!i0^+)^2\!-\!\xi_{\bf k}^2\!-\!\Delta_0^2}=\frac{E\tau_3\!+\!i\Delta_0\tau_2}{S^R(E)},  
\end{equation}
with $S^R(E)=\sqrt{(E+i0^+)^2-\Delta_0^2}$.

By defining $E_1=E+\Omega$ and $E_2=E+2\Omega$, from Eq.~(\ref{ELE1}), the equation of the first order of retarded Green function is written as 
\begin{eqnarray}
&&\!\!\!\!\!\!\!(E_1\tau_3\!+\!i\Delta_0\tau_2){\delta}g^{R(1)}(E)\!-\!{\delta}g^{R(1)}(E)(E\tau_3\!+\!i\Delta_0\tau_2)\nonumber\\
  &&\!\!\!\!\!\!\!\mbox{}=e{\bf A}_0\cdot{\bf v}_F\Pi^{R(0)}_3,\label{gr1}
\end{eqnarray}
from which one finds the first-order solution (refer to Appendix~\ref{BDCC}): 
\begin{equation}
 {\delta}g^{R(1)}(E)=(e{\bf A}_0\cdot{\bf v}_F)\frac{\tau_3-g^{R(0)}(E_1)\tau_3g^{R(0)}(E)}{S^R(E_1)+S^R(E)}. \label{sgr1}
\end{equation}
Here, $\Pi^{R(i)}_3=g^{R(i)}(E_1)\tau_3-\tau_3g^{R(i)}(E)$. Similarly, the equation of the second order of retarded Green function from Eq.~(\ref{ELE2}) reads
\begin{eqnarray}
&&\!\!\!\!\!\!\!(E_2\tau_3\!+\!i\Delta_0\tau_2){\delta}g^{R(2)}(E)\!-\!{\delta}g^{R(2)}(E)(E\tau_3\!+\!i\Delta_0\tau_2)\nonumber\\
  &&\!\!\!\!\!\!\!\mbox{}=e{\bf A}_0\!\cdot\!{\bf v}_F\Pi^{R(1)}_3\!+\!i\delta|\Delta|^{(2)}[g^{R(0)}(E_2)\tau_2\!-\!\tau_2g^{R(0)}(E)],~~~~~ \label{gr2}  
\end{eqnarray}
and gives the second-order solution (refer to Appendix~\ref{BDCC}):
\begin{eqnarray}
  &&\!\!\!\!\!\!\!{\delta}g^{R(2)}(E)=i\delta|\Delta|^{(2)}\frac{\tau_2\!-\!g^{R(0)}(E_2)\tau_3g^{R(0)}(E)}{S^R(E_2)+S^R(E)}\!+\!\frac{e{\bf A}_0\!\cdot\!{\bf v}_F}{E_2^2\!-\!E^2}\nonumber\\
    &&\!\!\!\!\!\!\!\mbox{}{\times}\big[S^R(E_2)g^{R(0)}(E_2)\Pi^{R(1)}_3\!-\!\Pi^{R(1)}_3S^R(E)g^{R(0)}(E)\big]. \label{sgr2}
\end{eqnarray}

Considering the response expansions, the normalization condition [Eq.~(\ref{nc})] for the retarded Green function is written as
\begin{eqnarray}
  &&[g^{R(0)}(E)]^2=1,\label{ncr0} \\
  &&g^{R(0)}(E_1){\delta}g^{R(1)}(E)\!+\!{\delta}g^{R(1)}(E)g^{R(0)}(E)=0,\label{ncr1} \\
  &&g^{R(0)}(E_2){\delta}g^{R(2)}(E)\!+\!{\delta}g^{R(2)}(E)g^{R(0)}(E)\nonumber\\
  &&\mbox{}\!+\!{\delta}g^{R(1)}(E_1){\delta}g^{R(1)}(E)=0.\label{ncr2}
\end{eqnarray}
The initial-state $g^{R(0)}(E)$ in Eq.~(\ref{gr0}) naturally satisfies Eq.~(\ref{ncr0}). Facilitated with Eq.~(\ref{ncr0}), correspondingly substituting the solutions in Eqs.~(\ref{sgr1}) and (\ref{sgr2}), one can easily demonstrate the normalization conditions in Eqs.~(\ref{ncr1}) and (\ref{ncr2}). Therefore, as the self-consistent crosscheck, the obtained solutions of the retarded Green function satisfy the normalization condition.

Further substituting Eq.~(\ref{gr0}) into Eqs.~(\ref{gr1}) and~(\ref{gr2}), the specific solution of the $\tau_2$ component of $\delta{g^{R(2)}}(E)$ is given by (refer to Appendix~\ref{BDCC})
\begin{eqnarray}\label{fsgr2}
 &&\!\!\!\!\!\!\! {\delta}g^{R(2)}_2(E)\!=\!i\delta|\Delta|^{(2)}\Big[\frac{4\Delta_0^2-(2\Omega)^2}{{\Gamma}^{(2)}(E)}\!+\!\frac{1}{S^R(E)}\Big]\nonumber\\
  &&\!\!\!\!\!\!\!\mbox{}+\frac{i(e{\bf A}_0\!\cdot\!{\bf v}_F)^2\Delta_0}{2\Omega^2}\Big[\frac{1}{S^R(E_2)}\!+\!\frac{1}{S^R(E)}\!-\!\frac{2}{S^R(E_1)}\Big], 
\end{eqnarray}
with $\Gamma^{(2)}(E)=2S^R(E_2)S^R(E)[S^R(E_2)\!+\!S^R(E)]$.

It is noted that the involved external optical frequency in the $\tau_2$ component of the second order of the retarded Green function in Eq.~(\ref{fsgr2}) is a continuous variable. To further consider the gap dynamics at nonzero temperature and eliminate the confusion of the auxiliary Matsubara frequency, we next derive the Keldysh Green function.

\subsection{Solution of Keldysh Green function}
\label{sec-KGF}

In this part, we derive the Keldysh Green function. We start with the normalization condition [Eq.~(\ref{nc})] for the Keldysh Green function:
\begin{equation}
g^R{\circ}g^{K}+g^K{\circ}g^A=0.  \label{ncgk}
\end{equation}
It is established that the Keldysh Green function can be written as a function of the retarded and advanced ones through a general relation\cite{QA1,Eilen1}:
\begin{equation}
g^K=g^R\circ{h}-h\circ{g^A}, \label{gkra}  
\end{equation}
where $h(t,t')$ denotes the distribution function. Substituting Eq.~(\ref{gkra}) to Eq.~(\ref{ncgk}), one finds that the normalization condition for the Keldysh Green function is immediately satisfied. Consequently, with the obtained retarded and hence advanced Green function in Sec.~\ref{SecGR}, to solve the Keldysh Green function, one only needs to solve the distribution function.  

In the previous work to derive the Ginzburg-Landau equation from Eilenberger equation within the Keldysh formalism\cite{Ba20}, the distribution function $h(t,t')$ is directly taken as the equilibrium one $\int\frac{dE}{2\pi}h(E)e^{-iE(t-t')}$ with the Fourier component written as
\begin{equation}\label{hE}
h(E)=\tanh\Big(\frac{\beta{E}}{2}\Big).  
\end{equation}
This treatment usually concerns the case near equilibrium or in strongly-interacting systems as applied in the transport theory of normal and superconducting metals\cite{QA1}, and has also been widely used in previous studies through the Eilenberger equation\cite{Eilen1,Ba8} and diffusive Usadel one\cite{Usadel,Usadel1}. In the present work, at clean limit, with a weak external excitation, we demonstrate Eq.~(\ref{hE}) by seriously taking account of the distribution function. 

Specifically, from the expansion of Green function matrices in Eq.~(\ref{RT}), the Keldysh Green function reads
\begin{eqnarray}
  g^K(t,t')&=&\int\frac{dE}{2\pi}e^{iE(t'-t)}[g^{K(0)}(E)\!+\!{\delta}g^{K(1)}(E)e^{-i{\Omega}t}\nonumber\\
  &&\mbox{}+{\delta}g^{K(2)}(E){e^{-2i{\Omega}t}}],~~~~~
\end{eqnarray}  
and with the general relation in Eq.~(\ref{gkra}), one has
\begin{eqnarray}
g^{K(0)}(E)\!\!\!&=&\!\!\!g^{R(0)}(E)h^{(0)}(E)-h^{(0)}(E)g^{A(0)}(E),\label{gk0}  \\
{\delta}g^{K(1)}(E)\!\!\!&=&\!\!\!g^{R(0)}(E_1){\delta}h^{(1)}(E)\!+\!{\delta}g^{R(1)}(E)h^{(0)}(E)\nonumber\\
  &&\mbox{}\!\!\!-\!h^{(0)}(E_1){\delta}g^{A(1)}(E)\!-\!{\delta}h^{(1)}(E)g^{A(0)}(E),~~~~~ \label{gk1}\\
{\delta}g^{K(2)}(E)\!\!\!&=&\!\!\!{\delta}g^{R(1)}(E_1){\delta}h^{(1)}(E)\!-\!{\delta}h^{(1)}(E_1){\delta}g^{A(1)}(E)\nonumber\\
  &&\mbox{}\!\!\!+\!g^{R(0)}(E_2){\delta}h^{(2)}(E)\!+\!{\delta}g^{R(2)}(E){h^{(0)}(E)}\nonumber\\
  &&\mbox{}\!\!\!-\!h^{(0)}(E_2){\delta}g^{A(2)}(E)\!-\!{\delta}h^{(2)}(E)g^{A(0)}(E).\label{gk2}
\end{eqnarray}
Here, ${\delta}h^{(1)}(E)$ and ${\delta}h^{(2)}(E)$ stand for the first- and second-order responses on the initial-state $h^{(0)}(E)$, respectively. 

According to Eq.~(\ref{ELE}), the Keldysh Green function satisfies the same equation as the retarded/advanced one. Consequently, by correspondingly replacing $g^{R(0)}$ and $\delta{g^{R(1)}}$ by $g^{K(0)}$ and $\delta{g^{K(1)}}$ in Eq.~(\ref{gr1}) and then substituting Eq.~(\ref{gk1}), one finds the equation of the first-order distribution function (refer to Appendix~\ref{CDCC}): 
\begin{eqnarray}
 &&\!\!\!\!\!\!\! (E_1\tau_3\!+\!i\Delta_0\tau_2)[g^{R(0)}(E_1){\delta}h^{(1)}(E)\!-\!{\delta}h^{(1)}(E)g^{A(0)}(E)]\nonumber\\
    &&\!\!\!\!\!\!\!\mbox{}-[g^{R(0)}(E_1){\delta}h^{(1)}(E)\!-\!{\delta}h^{(1)}(E)g^{A(0)}(E)](E\tau_3\!+\!i\Delta_0\tau_2)\nonumber\\
    &&\!\!\!\!\!\!\!\mbox{}=e{\bf A}_0\!\cdot\!{\bf v}_F[h^{(0)}(E_1)\!-\!h^{(0)}(E)][g^{R(0)}(E_1)\tau_3\!-\!\tau_3g^{A(0)}(E)].\nonumber\\ \label{eh1}
\end{eqnarray}
From above equation, the solution of the first-order distribution function reads (refer to Appendix~\ref{CDCC})
\begin{eqnarray}
  \delta{h}^{(1)}(E)&=&e{\bf A}_0\!\cdot\!{\bf v}_F\frac{h^{(0)}(E_1)\!-\!h^{(0)}(E)}{E_1^2\!-\!E^2}[(E_1\tau_3\!+\!i\Delta_0\tau_2)\tau_3\nonumber\\
    &&\mbox{}+\tau_3(E\tau_3\!+\!i\Delta_0\tau_2)]\nonumber\\
  &=&e{\bf A}_0\!\cdot\!{\bf v}_F\frac{h^{(0)}(E+\Omega)\!-\!h^{(0)}(E)}{\Omega}.\label{h1}
\end{eqnarray}  

Similarly, by correspondingly replacing $g^{R(0)}$ and $\delta{g^{R(i=1,2)}}$ by $g^{K(0)}$ and $\delta{g^{K(i=1,2)}}$ in Eq.~(\ref{gr2}), with Eqs.~(\ref{gk1}) and (\ref{gk2}) as well as the help of Eq.~(\ref{h1}), the equation of the second-order distribution function reads (refer to Appendix~\ref{CDCC})
\begin{eqnarray}
 &&\!\!\!\!\!\!\! (E_2\tau_3\!+\!i\Delta_0\tau_2)[g^{R(0)}(E_2){\delta}h^{(2)}(E)\!-\!{\delta}h^{(2)}(E)g^{A(0)}(E)]\nonumber\\
    &&\!\!\!\!\!\!\!\mbox{}-[g^{R(0)}(E_2){\delta}h^{(2)}(E)\!-\!{\delta}h^{(2)}(E)g^{A(0)}(E)](E\tau_3\!+\!i\Delta_0\tau_2)\nonumber\\
  &&\!\!\!\!\!\!\!\mbox{}=\!e{\bf A}_0\!\cdot\!{\bf v}_F[g^{R(0)}(E_2)\tau_3\!-\!\tau_3g^{A(0)}(E)][{\delta}h^{(1)}(E_1)\!-\!{\delta}h^{(1)}(E)]\nonumber\\
  &&\!\!\!\!\!\mbox{}+i\delta|\Delta|^{(2)}[h^{(0)}(E_2)\!-\!h^{(0)}(E)][g^{R(0)}(E_2)\tau_2\!-\!\tau_2g^{A(0)}(E)],\nonumber\\ \label{esh2}
\end{eqnarray}
from which the solution of the second-order distribution function is obtained as (refer to Appendix~\ref{CDCC})
\begin{eqnarray}
{\delta}h^{(2)}(E)&=&(e{\bf A}_0\!\cdot\!{\bf v}_F)\frac{{\delta}h^{(1)}(E_1)\!-\!{\delta}h^{(1)}(E)}{2\Omega}\nonumber\\
&&\mbox{}-\delta|\Delta|^{(2)}\frac{\Delta_0}{E\!+\!\Omega}\frac{h^{(0)}(E_2)\!-\!h^{(0)}(E)}{2\Omega}.\label{h2}
\end{eqnarray}  
Then, both the first- and second-order distribution functions are diagonal as they should be.

The initial-state distribution function can be obtained from the Hamiltonian in Eq.~(\ref{BdG}) in self-consistent consideration of the Higgs mode and vector potential, and is written as
\begin{eqnarray}\label{h0}
  h^{(0)}(E_{\bf k})&=&\tanh\Big\{\frac{\beta}{2}\big[\sqrt{\xi_{\bf k}^2\!+\!(\Delta_0\!+\!\delta|\Delta|)^2}\!-\!e{\bf A}\!\cdot\!{\bf v}_F\big]\Big\}\nonumber\\
  &\approx&\tanh\Big\{\frac{\beta}{2}\Big(E_{\bf k}+\frac{\Delta_0\delta|\Delta|}{E_{\bf k}}-e{\bf A}\!\cdot\!{\bf v}_F\Big)\Big\},
\end{eqnarray}
where $E_{\bf k}=\sqrt{\xi_{\bf k}^2+\Delta_0^2}$ denotes the Bogoliubov quasiparticle energy. Following the standard treatment of energy $E=E_{\bf k}$ as the previous work in
superconducting state\cite{Bo}, with the weak excitation, at low frequency ($\Omega<E=E_{\bf k}$), with Eqs.~(\ref{h1}) and (\ref{h2})-(\ref{h0}), the total distribution function in relative-frequency space reads
\begin{eqnarray}
  h(E)&=&h^{(0)}(E)+e^{-i\Omega{t}}\delta{h}^{(1)}(E)+e^{-2i\Omega{t}}\delta{h}^{(2)}(E) \nonumber\\
  &=&\Big[1\!+\!(e{\bf A}\!\cdot\!{\bf v}_F)\partial_{E}\!+\!\frac{(e{\bf A}\!\cdot\!{\bf v}_F)^2\partial_{E}^2}{2}\Big]h^{(0)}(E)\nonumber\\
  &&\mbox{}-\delta|\Delta|\frac{\Delta_0}{E}\partial_{E}h^{(0)}(E)\approx\tanh\Big(\frac{\beta{E}}{2}\Big).
\end{eqnarray}
Then, the drive effect of vector potential and Higgs-mode part in the initial-state distribution are exactly canceled by the first- and second-order distribution functions, leading to the widely applied distribution function [Eq.~(\ref{hE})] in the literature. \\

\subsection{Higgs-mode generation}

In this part, with the obtained distribution function in Eq.~(\ref{hE}) and the second-order retarded Green function in Eq.~(\ref{fsgr2}), from the gap equation [Eq.~(\ref{ELGE})], the second-order optical response of the Higgs mode at clean limit is determined by 
\begin{widetext}
\begin{eqnarray}
  \delta|\Delta|^{(2)}&=&-iU\int\frac{dE}{2\pi}\langle[h(E)\delta{g}^{R(2)}_2(E)-h(E_2)\delta{g}^{A(2)}_2(E)]\rangle_F=-iU\int\frac{dE}{2\pi}2h(E)\langle\delta{g}^{R(2)}_2(E)\rangle_F\nonumber\\
  &=&U\int{dE}2h(E)\Big\{\delta|\Delta|^{(2)}\Big[\frac{4\Delta_0^2-(2\Omega)^2}{{\Gamma}^{(2)}(E)}\!+\!\frac{1}{S^R(E)}\Big]+\frac{(e{A}_0v_F)^2\Delta_0}{6\Omega^2}\Big[\frac{1}{S^R(E_2)}\!+\!\frac{1}{S^R(E)}\!-\!\frac{2}{S^R(E_1)}\Big]\Big\}.
\end{eqnarray}
\end{widetext}
Consequently, one arrives at the equation of motion of the Higgs mode at clean limit:
\begin{equation}\label{EOMEE}
[4\Delta_0^2-(2\Omega)^2]\delta|\Delta|^{(2)}=-\frac{(e{A}_0v_F)^22\Delta_0}{3}\frac{\lambda_E}{\beta_E}, 
\end{equation}
similar to the one [Eq.~(\ref{GL-HME})] obtained from the Ginzburg-Landau Lagrangian. Here, through the standard contour integral, the amplitude-correlation coefficient reads
\begin{eqnarray}
  \beta_E&=&-\int\frac{dE}{2\pi}\frac{h(E)}{S^R(E_2)S^R(E)[S^R(E_2)\!+\!S^R(E)]}\nonumber\\
  &=&\sum_{n>0}\frac{1/(4\beta{i\Omega})}{p_n\!-\!i\Omega}\Big[\frac{1}{\sqrt{(p_n\!-\!2i\Omega)^2\!+\!\Delta_0^2}}\!-\!\frac{1}{\sqrt{p_n^2\!+\!\Delta_0^2}}\Big],\nonumber\\\label{betae}
\end{eqnarray}
and the essential response coefficient is given by 
\begin{eqnarray}
  \lambda_E&=&\!\!\!-\int\frac{dE}{2\pi}\frac{h(E)}{2\Omega^2}\Big[\frac{1}{S^R(E_2)}\!+\!\frac{1}{S^R(E)}\!-\!\frac{2}{S^R(E_1)}\Big]\nonumber\\
  &=&\!\!\!\frac{1}{2\beta\Omega^2}\sum_{n>0}\Big[\frac{2}{\sqrt{(p_n\!-\!i\Omega)^2\!+\!\Delta_0^2}}\!-\!\frac{1}{\sqrt{(p_n\!-\!2i\Omega)^2\!+\!\Delta_0^2}}\nonumber\\
    &&\mbox{}\!-\!\frac{1}{\sqrt{p_n^2\!+\!\Delta_0^2}}\Big].\label{lambdae}
\end{eqnarray}

It is noted that both amplitude-correlation coefficient $\beta_E$ [Eq.~(\ref{betae})] and response one $\lambda_E$ [Eq.~(\ref{lambdae})] derived here are exactly same as the ones obtained in the previous work\cite{Silaev} by Silaev through Eilenberger equation within Matsubara formalism. However, in Eqs.~(\ref{betae}) and (\ref{lambdae}), the fermionic Matsubara frequencies $ip_n$ arise from the singularities in the distribution function $h(E)$ during the standard contour integral in the complex plane, whereas the involved external optical frequency $\Omega$ within the Keldysh formalism is always a continuous variable, in contrast to the treatment of taking optical frequency as bosonic Matsubara frequency in Ref.~\onlinecite{Silaev}. As mentioned in the introduction, the treatment of taking optical frequency as bosonic Matsubara frequency leads to the vanishing response coefficient $\lambda_E$ (i.e., zero Higgs-mode generation) at all $\Omega\ne0$, strongly against the Ginzburg-Landau Lagrangian, whereas the prefactor $1/\Omega^2$ in $\lambda_E$ causes an undefined singularity at zero frequency, and hence, an unphysical discontinuity between cases at $\Omega=0$ and $\Omega\rightarrow0$. Actually, even for $\Omega\ne0$, from Eqs.~(\ref{betae})-(\ref{lambdae}), near $T_c$ with a weak gap,  one has
\begin{equation}
\beta_E\approx\frac{1}{4\beta\Omega^2}\sum_{n>0}\Big[\frac{2}{p_n\!-\!i\Omega}\!-\!\frac{1}{p_n\!-\!2i\Omega}\!-\!\frac{1}{{p_n}}\Big],\label{betac}
\end{equation}
and
\begin{equation}
 \lambda_E\approx\frac{1}{2\beta\Omega^2}\sum_{n>0}\Big[\frac{2}{p_n\!-\!i\Omega}\!-\!\frac{1}{p_n\!-\!2i\Omega}\!-\!\frac{1}{{p_n}}\Big].  \label{lambdac}
\end{equation}
In this circumstance, as $\beta_E=\lambda_E/2$, the Higgs-mode generation $\delta|\Delta|^{(2)}$, proportional to $\lambda_E/\beta_E$ from the equation of motion in Eq.~(\ref{EOMEE}), becomes undefined at Matsubara frequency $i\Omega_m$ which leads to $\lambda_E=\beta_E=0$. This directly poses a sharp challenge to the study in Ref.~\onlinecite{Silaev}. 

The derivation in the present study, which is performed in the Keldysh formalism and totally irrelevant to Matsubara space, naturally and analytically proves the continuous variable of the optical frequency in this situation. With the continuous optical frequency, near $T_c$, from Eqs.~(\ref{betac})-(\ref{lambdac}), one finds a finite Higgs-mode generation at all $\Omega$:
\begin{equation}
\delta|\Delta|^{(2)}=-\frac{(e{A}_0v_F)^2}{3}\frac{4\Delta_0}{[4\Delta_0^2-(2\Omega)^2]},  
\end{equation}
which exactly recovers the one [Eq.~(\ref{GL-HME})] derived from the Ginzburg-Landau Lagrangian. As for the regime with temperature far below $T_c$, with the continuous optical frequency, at low frequency ($\Omega<\Delta_0$), one finds the coefficient $\beta_E=\frac{1}{\beta}\sum_{n>0}({p_n^2+\Delta_0^2})^{-3/2}$ and in particular, a finite response coefficient:
\begin{equation}
 \lambda_E\approx\frac{1}{2\beta}\sum_{n>0}\partial_{p_n}^2\Big[\frac{1}{\sqrt{p_n^2\!+\!\Delta_0^2}}\Big],  
\end{equation}
implying a finite Higgs-mode generation at clean case. 

It is also noted that although the Eilenberger equation with the continuous optical frequency can recover the finite Higgs-mode generation at clean limit revealed by Ginzburg-Landau Lagrangian and gauge-invariant kinetic equation\cite{GIKE2,EPM}, this approach fails to derive the Higgs-mode damping by impurity scattering due to the generically incomplete scattering integral\cite{Silaev}. As proved in Ref.~\onlinecite{ESYW}, because of the quasiclassical approximation on $\tau_3$-Green function, the scattering integral in Eilenberger equation only involves the anisotropic part of the Green function that is related to the transport property, but generically drops out the isotropic one which determines the Higgs-mode lifetime. In this circumstance, the path-integral approach\cite{PYW} and gauge-invariant kinetic equation\cite{GIKE3} provide efficient and separate approaches to derive the induced damping of the Higgs mode by impurities, which agrees with the analysis through Heisenberg equation of motion as mentioned in the introduction and provides a possible origin for the experimentally observed broadening of the Higgs-mode resonance signal as well as the fast Higgs-mode damping after excitation\cite{GIKE3,PYW}.
\\

\section{summary}

In summary, we have resolved the current controversy in the literature that why the previous derivations at clean limit through the path-integral approach\cite{Cea1,Cea2,Cea3} and Eilenberger equation\cite{Silaev} within the Matsubara formalism failed to reach the Higgs-mode generation revealed by Ginzburg-Landau Lagrangian\cite{EPM} and gauge-invariant kinetic equation\cite{GIKE2,EPM}. The crucial treatment leading to this controversy lies at an artificial scheme within the Matsubara formalism that whether the involved external optical frequency $\Omega$ in the gap dynamics is taken as continuous variable or bosonic Matsubara frequency $i\Omega_m$. To resolve this confusion, we derive the effective action of superconducting gap near $T_c$ in the presence of the vector potential through the path-integral approach, and show that only by taking $\Omega$ as continuous variable within Matsubara formalism, one can achieve the fundamental Ginzburg-Landau superconducting Lagrangian in agreement with Landau phase-transition theory and symmetry analysis. In addition to this physical justification, we also perform the formulation of the gap dynamics within a separate and rigorous framework---Keldysh formalism, which is totally irrelevant to Matsubara space. By applying the Eilenberger equation in Keldysh space to calculate the second-order response of the Higgs mode, it is analytically proved that the involved optical frequency is always a continuous variable, leading to finite response coefficient at clean limit. 

Consequently, the present study confirms the unified conclusion, i.e., a finite Higgs-mode generation at clean limit in the second-order response of superconductors from three different microscopic approaches (including the gauge-invariant kinetic equation, Eilenberger equation and path-integral approach) as well as from Ginzburg-Landau Lagrangian, and can therefore help understanding the experimental findings of the observed Higgs-mode excitation\cite{NL7,NL8,NL9,NL10,NL11,DHM2,DHM3}.

\begin{acknowledgments}
The authors acknowledge financial support from
the National Natural Science Foundation of 
China under Grants No.\ 11334014 and No.\ 61411136001.  
\end{acknowledgments}

\begin{widetext}

\begin{appendix}
    
\section{Derivation of correlation coefficients}
\label{ADCC}

In this part, we present the specific expressions of the related correlation coefficients in the superconducting Lagrangian in Eq.~(\ref{LG1}). Firstly, as the Green function $G_0(p)=\frac{p_0+\xi_{\bf k}\tau_3}{p_0^2-\xi_{\bf k}^2}$ only consists of the $\tau_0$ and $\tau_3$ components, from Eqs.~(\ref{chii})-(\ref{chiijkl}), one immediately finds $\chi_1=\chi_{13}=\chi_{111}=\chi_{100}=\chi_{1113}=0$. Moreover, within the Matsubara formalism $[p=(ip_n,{\bf k})]$, one has
\begin{eqnarray}
   \chi_{113}&=&\sum_p{\rm Tr}[G_0(p\!+\!2q)\tau_1G_0(p\!+\!q)\tau_1G_0(p)\tau_3]=\sum_{p}\frac{2\xi_{\bf k}[(ip_n\!+\!2\Omega)(ip_n\!+\!\Omega)\!-\!\xi_{\bf k}^2\!-\!ip_n\Omega]}{[(ip_n\!+\!2\Omega)^2\!-\!\xi_{\bf k}^2][(ip_n\!+\!\Omega)^2\!-\!\xi_{\bf k}^2][(ip_n)^2\!-\!\xi_{\bf k}^2]}=0,~~~~~\\
  \chi_p&=&\frac{1}{2}\sum_p{\rm Tr}[G_0(p\!+\!q)\tau_1G_0(p)\tau_1]\!+\!\frac{1}{U}=\sum_p\frac{\!(ip_n\!+\!\Omega)^2\!+\!(ip_n)^2\!-\!(ip_n\!+\!\Omega\!-\!ip_n)^2\!-\!2\xi_{\bf k}^2}{2[(ip_n\!+\!\Omega)^2\!-\!\xi_{\bf k}^2][(ip_n)^2\!-\!\xi_{\bf k}^2]}\!+\!\frac{1}{U}\nonumber\\
  &\approx&-\frac{\Omega^2}{2}\sum_p\frac{1}{[(ip_n\!+\!\Omega)^2\!-\!\xi_{\bf k}^2][(ip_n)^2\!-\!\xi_{\bf k}^2]}\!+\!\sum_{\bf k}\frac{f(\xi_{\bf k})-f(-\xi_{\bf k})}{2\xi_{\bf k}}\!+\!\frac{1}{U}\nonumber\\
  &=&-\frac{\Omega^2}{2}\sum_p\frac{1}{[(ip_n\!+\!\Omega)^2\!-\!\xi_{\bf k}^2][(ip_n)^2\!-\!\xi_{\bf k}^2]}\!+\!D\int^{\omega_D}_{-\omega_D}d\xi_{\bf k}\frac{\tanh({\beta_c\xi_{\bf k}}/{2})\!-\!\tanh({\beta\xi_{\bf k}}/{2})}{2\xi_{\bf k}},\\
  \chi_{1111}&=&\sum_p{\rm Tr}[G_0(p\!+\!q)\tau_1G_0(p)\tau_1G_0(p\!+\!q)\tau_1G(p)\tau_1]=\sum_p\frac{2}{(ip_n\!+\!\Omega\!-\!\xi_{\bf k})^2(ip_n\!+\!\xi_{\bf k})^2},\nonumber\\
  \chi_{1010}&=&\sum_p{\rm Tr}[G_0(p\!+\!q)\tau_1G_0(p)\tau_0G_0(p\!+\!q)\tau_1G(p)\tau_0]=\sum_p\frac{2}{[(ip_n\!+\!\Omega)^2\!-\!\xi^2_{\bf k}][(ip_n)^2\!-\!\xi^2_{\bf k}]},\\ 
  \chi_{1100}\!+\!\chi_{0110}&=&\sum_p\Big[\frac{2}{(ip_n\!+\!\Omega\!-\!\xi_{\bf k})^2[(ip_n)^2\!-\!\xi^2_{\bf k}]}+\frac{2}{[(ip_n\!+\!\Omega)^2\!-\!\xi^2_{\bf k}](ip_n\!-\!\xi_{\bf k})^2}\Big].
\end{eqnarray}
Here, we have used the gap equation $\frac{1}{U}=D\int^{\omega_D}_{-\omega_D}d\xi_{\bf k}\frac{\tanh({\beta_c\xi_{\bf k}}/{2})}{2\xi_{\bf k}}$ at the critical temperature in the BCS theory\cite{G1}. It is noted that $\chi_{113}$ vanishes as the consequence of the particle-hole symmetry, which eliminates the terms with the odd order of $\xi_{\bf k}$ in the summation of ${\bf k}$.

Then, further using the facts:
\begin{eqnarray} 
&&\sum_p\frac{1}{[(ip_n\!+\!\Omega)^2\!-\!\xi^2_{\bf k}][(ip_n)^2\!-\!\xi^2_{\bf k}]}=\frac{D}{\beta}\sum_{ip_{n>0},\eta=\pm}\int{d\xi_{\bf k}}\frac{1}{[(ip_n\!+\!\eta\Omega)^2\!-\!\xi_{\bf k}^2][(ip_n)^2\!-\!\xi_{\bf k}^2]}\nonumber\\
  &&\mbox{}=\frac{2\pi{i}D}{\beta\Omega}\sum_{ip_{n>0},\eta=\pm}\Big[\frac{1}{(2ip_n\!+\!2\Omega)(2ip_n\!+\!\Omega)}\!-\!\frac{1}{2ip_n(2ip_n\!+\!\Omega)}\Big]=\frac{\pi{i}D}{\beta\Omega^2}\sum_{ip_{n>0},\eta=\pm}\Big[\frac{4}{2ip_n\!+\!\Omega}\!-\!\frac{1}{ip_n}\!-\!\frac{1}{ip_n\!+\!\Omega}\Big],\\
  && \sum_p\frac{2}{(ip_n\!+\!\Omega\!-\!\xi_{\bf k})^2(ip_n\!+\!\xi_{\bf k})^2}=\frac{D}{\beta}\sum_{ip_{n>0},\eta=\pm}\int{d\xi_{\bf k}}\frac{2}{(ip_n\!+\!\eta\Omega\!-\!\xi_{\bf k})^2(ip_n\!+\!\xi_{\bf k})^2}=-\sum_{ip_{n>0},\eta=\pm}\frac{8\pi{i}D/\beta}{(2ip_n\!+\!\eta\Omega)^3},\\
  &&\sum_p\Big[\frac{2}{(ip_n\!+\!\Omega\!-\!\xi_{\bf k})^2[(ip_n)^2\!-\!\xi^2_{\bf k}]}+\frac{2}{[(ip_n\!+\!\Omega)^2\!-\!\xi^2_{\bf k}](ip_n\!-\!\xi_{\bf k})^2}\Big]\nonumber\\
  &&\mbox{}=\frac{D}{\beta}\sum_{ip_{n>0},\eta=\pm}\int{d\xi_{\bf k}}\Big[\frac{2}{(ip_n\!+\!\eta\Omega\!-\!\xi_{\bf k})^2[(ip_n)^2\!-\!\xi^2_{\bf k}]}+\frac{2}{[(ip_n\!+\!\eta\Omega)^2\!-\!\xi^2_{\bf k}](ip_n\!-\!\xi_{\bf k})^2}\Big]\nonumber\\
  &&\mbox{}=\frac{2\pi{i}D}{\beta\Omega^2}\sum_{ip_{n>0},\eta=\pm}\Big[\frac{4}{2ip_n\!+\!\eta\Omega}\!-\!\frac{1}{ip_n}\!-\!\frac{1}{ip_n\!+\!\eta\Omega}\Big],
\end{eqnarray}
the Landau parameters $\beta_p$ [Eq.~(\ref{betapp})], $\lambda_p$ [Eq.~(\ref{f1010})], $\alpha_p$  [Eq.~(\ref{apf})] and $\gamma_p$ [Eq.~(\ref{fgp})] are derived.

\section{Derivation of retarded Green function from Eilenberger equation}
\label{BDCC}

In this part, we present the derivation of the retarded Green function from the Eilenberger equation. With the initial state of the retarded Green function in Eq.~(\ref{gr0}), the equation of the first order of retarded Green function in Eq.~(\ref{gr1}) is re-written as
\begin{equation}
S^R(E_1)g^{R(0)}(E_1){\delta}g^{R(1)}(E)\!-\!{\delta}g^{R(1)}(E)S^R(E)g^{R(0)}(E)=e{\bf A}_0\cdot{\bf v}_F\Pi^{R(0)}_3.
\end{equation}
From above equation, considering the normalization condition of $g^{R(0)}(E)$ in Eq.~(\ref{ncr0}), one has
\begin{eqnarray}
  [S^R(E_1)]^2{\delta}g^{R(1)}(E)\!-\!S^R(E_1)S^R(E)g^{R(0)}(E_1){\delta}g^{R(1)}(E)g^{R(0)}(E)=S^R(E_1)e{\bf A}_0\cdot{\bf v}_F[\tau_3\!-\!g^{R(0)}(E_1)\tau_3g^{R(0)}(E)],\label{B2}\\
  S^R(E)S^R(E_1)g^{R(0)}(E_1){\delta}g^{R(1)}(E)g^{R(0)}(E)\!-\![S^R(E)]^2{\delta}g^{R(1)}(E)=S^R(E)e{\bf A}_0\cdot{\bf v}_F[g^{R(0)}(E_1)\tau_3g^{R(0)}(E)\!-\!\tau_3].\label{B3}
\end{eqnarray}
Then, the solution of ${\delta}g^{R(1)}(E)$ in Eq.~(\ref{sgr1}) can be easily obtained by adding Eqs.~(\ref{B2}) and~(\ref{B3}). Moreover, substituting Eq.~(\ref{gr0}) into Eq.~(\ref{sgr1}), the specific expression of ${\delta}g^{R(1)}(E)$ is given by
\begin{equation}\label{sssgr1}
  {\delta}g^{R(1)}(E)=(e{\bf A}_0\cdot{\bf v}_F)\frac{\tau_3-[S^R(E_1)S^R(E)]^{-1}[E_1E\tau_3+i\Delta_0(E_0+E_1)\tau_2+\Delta_0^2\tau_3]}{S^R(E_1)+S^R(E)}.
\end{equation}

Similarly, the equation of the second order of retarded Green function in Eq.~(\ref{gr2}) is re-written as
\begin{equation}
S^R(E_2)g^{R(0)}(E_2){\delta}g^{R(2)}(E)\!-\!{\delta}g^{R(2)}(E)S^R(E)g^{R(0)}(E)=e{\bf A}_0\!\cdot\!{\bf v}_F\Pi^{R(1)}_3\!+\!i\delta|\Delta|^{(2)}[g^{R(0)}(E_2)\tau_2\!-\!\tau_2g^{R(0)}(E)],
\end{equation}
and using the normalization condition of $g^{R(0)}(E)$ in Eq.~(\ref{ncr0}), one easily gets the solution of ${\delta}g^{R(1)}(E)$ in Eq.~(\ref{sgr2}). Substituting Eqs.~(\ref{gr0}) and~(\ref{sssgr1}) into Eq.~(\ref{sgr2}), the specific expression of the $\tau_2$ component of ${\delta}g^{R(2)}(E)$ reads
\begin{eqnarray}
  {\delta}g^{R(2)}_2(E)&=&{i(e{\bf A}_0\!\cdot\!{\bf v}_F)^2\Delta_0}\Big\{\frac{(EE_2\!+\!E_1E\!+E_1E_2\!+\!\Delta_0^2)[S^R(E_1)\!+\!S^R(E_2)\!+\!S^R(E)]}{S^R(E_1)S^R(E_2)S^R(E)}-1\Big\}\frac{1}{[S^R(E)\!+\!S^R(E_2)]} \nonumber\\
  &&\mbox{}\times\frac{1}{[S^R(E_1)\!+\!S^R(E_2)][S^R(E_1)\!+\!S^R(E)]}\!+\!i\delta|\Delta|^{(2)}\frac{2\Delta_0^2\!+\!2EE_2\!+\!2S^R(E)S^R(E_2)}{\Gamma^{(2)}(E)}.
\end{eqnarray}

Further considering
\begin{equation}
\frac{[S^R(E_1)\!+\!S^R(E_2)\!+\!S^R(E)][S^R(E)\!-\!S^R(E_1)][S^R(E_1)\!-\!S^R(E_2)][S^R(E)\!-\!S^R(E_2)]}{S^R(E_1)S^R(E_2)S^R(E)}=\frac{E_1^2\!-\!E_2^2}{S^R(E)}\!+\!\frac{E_2^2\!-\!E^2}{S^R(E_1)}\!+\!\frac{E^2\!-\!E_1^2}{S^R(E_2)},~
\end{equation}
one has
\begin{eqnarray}
  {\delta}g^{R(2)}_2(E)&=&\frac{i(e{\bf A}_0\!\cdot\!{\bf v}_F)^2\Delta_0}{(E^2_1-E^2)(E_2^2-E^2)(E_1^2-E_2^2)}\Big\{{(EE_2\!+\!E_1E\!+E_1E_2\!+\!\Delta_0^2)}{}\Big[\frac{E_1^2\!-\!E_2^2}{S^R(E)}\!+\!\frac{E_2^2\!-\!E^2}{S^R(E_1)}\!+\!\frac{E^2\!-\!E_1^2}{S^R(E_2)}\Big]\nonumber\\
  &&\mbox{}-{[S^R(E)\!-\!S^R(E_2)][S^R(E_1)\!-\!S^R(E_2)][S^R(E)\!-\!S^R(E_1)]}\Big\}+i\delta|\Delta|^{(2)}\frac{2\Delta_0^2\!+\!2EE_2\!+\!2S^R(E)S^R(E_2)}{\Gamma^{(2)}(E)}\nonumber\\
  &=&\frac{i(e{\bf A}_0\!\cdot\!{\bf v}_F)^2\Delta_0}{(E^2_1\!-\!E^2)(E_2^2\!-\!E^2)(E_1^2\!-\!E_2^2)}\Big\{(E_1^2\!-\!E_2^2)\frac{(E\!+\!E_2)(E_1\!+\!E)\!-\![S^R(E)]^2}{S^R(E)}\!+\!\frac{(E\!+\!E_1)(E_1\!+\!E_2)\!-\![S^R(E_1)]^2}{S^R(E_1)}\nonumber\\
  &&\mbox{}\times(E_2^2\!-\!E^2)\!+\!(E^2\!-\!E_1^2)\frac{(E\!+\!E_2)(E_2\!+\!E_1)\!-\![S^R(E_2)]^2}{S^R(E_2)}\!+\!S^R(E)\big\{[S^R(E_1)]^2\!-\![S^R(E_2)]^2\big\}\!+\!\big\{[S^R(E_2)]^2\nonumber\\
    &&\mbox{}-[S^R(E)]^2\big\}S^R(E_1)\!+\!\!S^R(E_2)\big\{[S^R(E)]^2\!-\![S^R(E_1)]^2\big\}\Big\}\!+\!i\delta|\Delta|^{(2)}\frac{4\Delta_0^2\!-\!(E\!-\!E_2)^2\!+\![S^R(E)\!+\!S^R(E_2)]^2}{\Gamma^{(2)}(E)}\nonumber\\
    &=&\frac{i(e{\bf A}_0\!\cdot\!{\bf v}_F)^2\Delta_0}{2\Omega^2}\Big[\frac{1}{S^R(E_2)}\!+\!\frac{1}{S^R(E)}\!-\!\frac{2}{S^R(E_1)}\Big]+i\delta|\Delta|^{(2)}\Big[\frac{4\Delta_0^2-(2\Omega)^2}{{\Gamma}^{(2)}(E)}\!+\!\frac{1}{S^R(E)}\Big].
\end{eqnarray}
Then, Eq.~(\ref{fsgr2}) is derived.

\section{Derivation of distribution function from Eilenberger equation}
\label{CDCC}

In this part, we present the derivation of the distribution function from the Eilenberger equation. As mentioned in Sec.~\ref{sec-KGF}, according to Eq.~(\ref{ELE}), the Keldysh Green function satisfies the same equation as the retarded/advanced one. Therefore, the equation of the first order of the Keldysh Green function reads
\begin{equation}
(E_1\tau_3\!+\!i\Delta_0\tau_2){\delta}g^{K(1)}(E)\!-\!{\delta}g^{K(1)}(E)(E\tau_3\!+\!i\Delta_0\tau_2)=e{\bf A}_0\cdot{\bf v}_F\Pi^{K(0)}_3.  
\end{equation}
Substituting $g^{K(0)}$ [Eq.~(\ref{gk0})] and ${\delta}g^{K(1)}(E)$ [Eq.~(\ref{gk1})], the above equation becomes
\begin{eqnarray}
 &&\!\!\!\!\!\!\!(E_1\tau_3\!+\!i\Delta_0\tau_2)[g^{R(0)}(E_1){\delta}h^{(1)}(E)\!-\!{\delta}h^{(1)}(E)g^{A(0)}(E)]\!-\![g^{R(0)}(E_1){\delta}h^{(1)}(E)\!-\!{\delta}h^{(1)}(E)g^{A(0)}(E)](E\tau_3\!+\!i\Delta_0\tau_2)\!=\!h^{(0)}(E)\nonumber\\
  &&\!\!\!\!\!\!\!\mbox{}\times[{\delta}g^{R(1)}(E)(E\tau_3\!+\!i\Delta_0\tau_2)\!-\!(E_1\tau_3\!+\!i\Delta_0\tau_2){\delta}g^{R(1)}(E)]\!-\!h^{(0)}(E_1)[{\delta}g^{A(1)}(E)(E\tau_3\!+\!i\Delta_0\tau_2)\!-\!(E_1\tau_3\!+\!i\Delta_0\tau_2){\delta}g^{A(1)}(E)]\nonumber\\
  &&\!\!\!\!\!\!\!\mbox{}-e{\bf A}_0\!\cdot\!{\bf v}_F[h^{(0)}(E)\tau_3g^{R(0)}(E)\!-\!g^{R(0)}(E_1)h^{(0)}(E_1)\tau_3\!-\!h^{(0)}(E)\tau_3g^{A(0)}(E)\!+\!g^{A(0)}(E_1)h^{(0)}(E_1)\tau_3].~~~~
\end{eqnarray}
Then, facilitated with the equations of the first order of the retarded [Eq.~(\ref{gr1})] and advanced Green functions, one arrives at the equation of the first order of the distribution function in Eq.~(\ref{eh1}). By first multiplying Eq.~(\ref{eh1}) by $(E_1\tau_3+i\Delta_0\tau_2)$ from the left side and $(E\tau_3+i\Delta_0\tau_2)$ from the right side respectively and adding the obtained two equations afterwards, one has  
\begin{eqnarray}
  &&[(E_1^2\!+\!\Delta_0^2)\!-\!(E^2\!+\!\Delta_0^2)][g^{R(0)}(E_1){\delta}h^{(1)}(E)\!-\!{\delta}h^{(1)}(E)g^{A(0)}(E)]=e{\bf A}_0\!\cdot\!{\bf v}_F[h^{(0)}(E_1)\!-\!h^{(0)}(E)]\nonumber\\
  &&\mbox{}\times\big\{g^{R(0)}(E_1)[(E_1\tau_3\!+\!i\Delta_0\tau_2)\tau_3\!+\!\tau_3(E\tau_3\!+\!i\Delta_0\tau_2)]\!-\![(E_1\tau_3\!+\!i\Delta_0\tau_2)\tau_3\!+\!\tau_3(E\tau_3\!+\!i\Delta_0\tau_2)]g^{A(0)}(E)\big\},
\end{eqnarray}
from which the solution of the first order of the distribution function reads
\begin{equation}
{\delta}h^{(1)}(E)=e{\bf A}_0\!\cdot\!{\bf v}_F[h^{(0)}(E_1)\!-\!h^{(0)}(E)]\frac{[(E_1\tau_3\!+\!i\Delta_0\tau_2)\tau_3\!+\!\tau_3(E\tau_3\!+\!i\Delta_0\tau_2)]}{(E_1^2\!+\!\Delta_0^2)\!-\!(E^2\!+\!\Delta_0^2)},  
\end{equation}
and then, Eq.~(\ref{h1}) is derived. 

Similarly, the equation of the second order of the Keldysh Green function reads
\begin{equation}
(E_2\tau_3\!+\!i\Delta_0\tau_2){\delta}g^{K(2)}(E)\!-\!{\delta}g^{K(2)}(E)(E\tau_3\!+\!i\Delta_0\tau_2)=e{\bf A}_0\!\cdot\!{\bf v}_F\Pi^{K(1)}_{3}\!+\!i\delta|\Delta|^{(2)}[g^{K(0)}(E_2)\tau_2\!-\!\tau_2g^{K(0)}(E)].  
\end{equation}
Substituting $g^{K(0)}$ [Eq.~(\ref{gk0})] and ${\delta}g^{K(1)}(E)$ [Eq.~(\ref{gk1})] as well as ${\delta}g^{K(2)}(E)$ [Eq.~(\ref{gk2})], the above equation becomes
\begin{eqnarray}
  &&\!\!\!\!\!\!\!(E_2\tau_3\!+\!i\Delta_0\tau_2)[g^{R(0)}(E_2){\delta}h^{(2)}(E)\!-\!{\delta}h^{(2)}(E)g^{A(0)}(E)]\!-\![g^{R(0)}(E_2){\delta}h^{(2)}(E)\!-\!{\delta}h^{(2)}(E)g^{A(0)}(E)](E\tau_3\!+\!i\Delta_0\tau_2)\nonumber\\
  &&\!\!\!\!\!\!\!=[{\delta}g^{R(1)}(E_1){\delta}h^{(1)}(E)\!-\!{\delta}h^{(1)}(E_1){\delta}g^{A(1)}(E)](E\tau_3\!+\!i\Delta_0\tau_2)\!-\!(E_2\tau_3\!+\!i\Delta_0\tau_2)[{\delta}g^{R(1)}(E_1){\delta}h^{(1)}(E)\!-\!{\delta}h^{(1)}(E_1){\delta}g^{A(1)}(E)]\nonumber\\
  &&\!\!\!\!\!\!\!\mbox{}+h^{(0)}(E)[{\delta}g^{R(2)}(E)(E\tau_3\!+\!i\Delta_0\tau_2)\!-\!(E_2\tau_3\!+\!i\Delta_0\tau_2){\delta}g^{R(2)}(E)]\!-\![{\delta}g^{A(2)}(E)(E\tau_3\!+\!i\Delta_0\tau_2)\!-\!(E_2\tau_3\!+\!i\Delta_0\tau_2){\delta}g^{A(2)}(E)]\nonumber\\ 
  &&\!\!\!\!\!\!\mbox{}{\times}h^{(0)}(E_2)\!+\!e{\bf A}_0\!\cdot\!{\bf v}_F[{\delta}g^{R(1)}(E_1)h^{(0)}(E_1)\tau_3\!-\!\tau_3{\delta}g^{R(1)}(E)h^{(0)}(E)\!+\!g^{R(0)}(E_2){\delta}h^{(1)}(E_1)\tau_3\!-\!\tau_3g^{R(0)}(E_1){\delta}h^{(1)}(E)]\nonumber\\
   &&\!\!\!\!\!\!\!\mbox{}-e{\bf A}_0\!\cdot\!{\bf v}_F[h^{(0)}(E_2){\delta}g^{A(1)}(E_1)\tau_3\!-\!\tau_3h^{(0)}(E_1){\delta}g^{A(1)}(E)\!+\!{\delta}h^{(1)}(E_1)g^{A(0)}(E_1)\tau_3\!-\!\tau_3{\delta}h^{(1)}(E)g^{A(0)}(E)]\nonumber\\
   &&\!\!\!\!\!\!\!\mbox{}-i\delta|\Delta|^{(2)}[\tau_2g^{R(0)}(E)h^{(0)}(E)\!-\!g^{R(0)}(E_2)h^{(0)}(E_2)\tau_2\!-\!\tau_2g^{A(0)}(E)h^{(0)}(E)\!+\!g^{A(0)}(E_2)h^{(0)}(E_2)\tau_2].
\end{eqnarray}
Facilitated with the equations of the second order of the retarded [Eq.~(\ref{gr2})] and advanced Green functions, one has
\begin{eqnarray}
  &&\!\!\!\!\!\!\!(E_2\tau_3\!+\!i\Delta_0\tau_2)[g^{R(0)}(E_2){\delta}h^{(2)}(E)\!-\!{\delta}h^{(2)}(E)g^{A(0)}(E)]\!-\![g^{R(0)}(E_2){\delta}h^{(2)}(E)\!-\!{\delta}h^{(2)}(E)g^{A(0)}(E)](E\tau_3\!+\!i\Delta_0\tau_2)\nonumber\\
  &&\!\!\!\!\!\!\!=[{\delta}g^{R(1)}(E_1){\delta}h^{(1)}(E)\!-\!{\delta}h^{(1)}(E_1){\delta}g^{A(1)}(E)](E\tau_3\!+\!i\Delta_0\tau_2)\!-\!(E_2\tau_3\!+\!i\Delta_0\tau_2)[{\delta}g^{R(1)}(E_1){\delta}h^{(1)}(E)\!-\!{\delta}h^{(1)}(E_1){\delta}g^{A(1)}(E)]\nonumber\\
  &&\!\!\!\!\!\!\!\mbox{}+e{\bf A}_0\!\cdot\!{\bf v}_F[h^{(0)}(E_1)-h^{(0)}(E)]{\delta}g^{R(1)}(E_1)\tau_3\!+\!e{\bf A}_0\!\cdot\!{\bf v}_F[g^{R(0)}(E_2){\delta}h^{(1)}(E_1)\tau_3\!-\!\tau_3g^{R(0)}(E_1){\delta}h^{(1)}(E)]\nonumber\\
   &&\!\!\!\!\!\!\!\mbox{}-e{\bf A}_0\!\cdot\!{\bf v}_F[h^{(0)}(E_2)\!-\!h^{(0)}(E_1)]\tau_3{\delta}g^{A(1)}(E)\!-\!e{\bf A}_0\!\cdot\!{\bf v}_F[{\delta}h^{(1)}(E_1)g^{A(0)}(E_1)\tau_3\!-\!\tau_3{\delta}h^{(1)}(E)g^{A(0)}(E)]\nonumber\\
  &&\!\!\!\!\!\!\!\mbox{}+i\delta|\Delta|^{(2)}[h^{(0)}(E_2)\!-\!h^{(0)}(E)][g^{R(0)}(E_2)\tau_2\!-\!\tau_2g^{A(0)}(E)].
\end{eqnarray}
Further using Eq.~(\ref{h1}) to replace $e{\bf A}_0\!\cdot\!{\bf v}_F[h^{(0)}(E_1)-h^{(0)}(E)]$ with $\Omega\delta{h^{1}}(E)$, the above equation is simplified as 
\begin{eqnarray}
  &&\!\!\!\!\!\!\!(E_2\tau_3\!+\!i\Delta_0\tau_2)[g^{R(0)}(E_2){\delta}h^{(2)}(E)\!-\!{\delta}h^{(2)}(E)g^{A(0)}(E)]\!-\![g^{R(0)}(E_2){\delta}h^{(2)}(E)\!-\!{\delta}h^{(2)}(E)g^{A(0)}(E)](E\tau_3\!+\!i\Delta_0\tau_2)\nonumber\\
  &&\!\!\!\!\!\!\!={\delta}h^{(1)}(E)[{\delta}g^{R(1)}(E_1)(E_1\tau_3\!+\!i\Delta_0\tau_2)\!-\!(E_2\tau_3\!+\!i\Delta_0\tau_2){\delta}g^{R(1)}(E_1)\!-\!e{\bf A}_0\!\cdot\!{\bf v}_F\tau_3g^{R(0)}(E_1)]\nonumber\\
  &&\!\!\!\!\!\!\!\mbox{}-{\delta}h^{(1)}(E_1)[{\delta}g^{A(1)}(E_1)(E_1\tau_3\!+\!i\Delta_0\tau_2)\!-\!(E_2\tau_3\!+\!i\Delta_0\tau_2){\delta}g^{A(1)}(E_1)\!+\!e{\bf A}_0\!\cdot\!{\bf v}_Fg^{A(0)}(E_1)\tau_3]\nonumber\\
  &&\!\!\!\!\!\!\!\mbox{}+e{\bf A}_0\!\cdot\!{\bf v}_F[g^{R(0)}(E_2){\delta}h^{(1)}(E_1)\tau_3\!+\!\tau_3{\delta}h^{(1)}(E)g^{A(0)}(E)]\!+\!i\delta|\Delta|^{(2)}[h^{(0)}(E_2)\!-\!h^{(0)}(E)][g^{R(0)}(E_2)\tau_2\!-\!\tau_2g^{A(0)}(E)].~~~~
\end{eqnarray}
Consequently, substituting equations of the first order of the retarded [Eq.~(\ref{gr1})] and advanced Green functions to above equation, one arrives at the equation of the second order of the distribution function in Eq.~(\ref{esh2}). By first multiplying Eq.~(\ref{eh1}) by $(E_2\tau_3+i\Delta_0\tau_2)$ from the left side and $(E\tau_3+i\Delta_0\tau_2)$ from the right side respectively and adding the obtained two equations afterwards, one has  
\begin{eqnarray}
  &&\!\!\!\!\!\![g^{R(0)}(E_2){\delta}h^{(2)}(E)\!-\!{\delta}h^{(2)}(E)g^{A(0)}(E)]\!=\!\frac{e{\bf A}_0\!\cdot\!{\bf v}_F[{\delta}h^{(1)}(E_1)\!-\!{\delta}h^{(1)}(E)]}{E_2^2\!-\!E^2}\big\{g^{R(0)}(E_2)[(E_2\tau_3\!+\!i\Delta_0\tau_2)\tau_3\!+\!\tau_3(E\tau_3\!+\!i\Delta_0\tau_2)]\nonumber\\
  &&\!\!\!\!\!\!\mbox{}-[(E_2\tau_3\!+\!i\Delta_0\tau_2)\tau_3\!+\!\tau_3(E\tau_3\!+\!i\Delta_0\tau_2)]g^{A(0)}(E)\big\}\!+\!i\delta|\Delta|^{(2)}\frac{h^{(0)}(E_2)\!-\!h^{(0)}(E)}{E_2^2\!-\!E^2}\big\{g^{R(0)}(E_2)[(E_2\tau_3\!+\!i\Delta_0\tau_2)\tau_2\!+\!\tau_2(E\tau_3\nonumber\\
  &&\!\!\!\!\!\!\mbox{}+i\Delta_0\tau_2)]\!-\![(E_2\tau_3\!+\!i\Delta_0\tau_2)\tau_2\!+\!\tau_2(E\tau_3\!+\!i\Delta_0\tau_2)]g^{A(0)}(E)\big\},
\end{eqnarray}
from which the solution of the first order of the distribution function reads
\begin{eqnarray}
  {\delta}h^{(2)}(E)&=&e{\bf A}_0\!\cdot\!{\bf v}_F\frac{[{\delta}h^{(1)}(E_1)\!-\!{\delta}h^{(1)}(E)][(E_2\tau_3\!+\!i\Delta_0\tau_2)\tau_3\!+\!\tau_3(E\tau_3\!+\!i\Delta_0\tau_2)]}{E_2^2\!-\!E^2}\nonumber\\
  &&+i\delta|\Delta|^{(2)}\frac{[h^{(0)}(E_2)\!-\!h^{(0)}(E)][(E_2\tau_3\!+\!i\Delta_0\tau_2)\tau_2\!+\!\tau_2(E\tau_3\!+\!i\Delta_0\tau_2)]}{E_2^2\!-\!E^2}\nonumber\\
  &=&e{\bf A}_0\!\cdot\!{\bf v}_F\frac{{\delta}h^{(1)}(E_1)\!-\!{\delta}h^{(1)}(E)}{E_2-E}\!+\!i\delta|\Delta|^{(2)}\frac{[h^{(0)}(E_2)\!-\!h^{(0)}(E)](2i\Delta_0\!-\!2i\Omega{\tau_1})}{E_2^2\!-\!E^2}\nonumber\\
   &\approx&e{\bf A}_0\!\cdot\!{\bf v}_F\frac{{\delta}h^{(1)}(E_1)\!-\!{\delta}h^{(1)}(E)}{E_2-E}\!+\!i\delta|\Delta|^{(2)}2i\Delta_0\frac{h^{(0)}(E_2)\!-\!h^{(0)}(E)}{E_2^2\!-\!E^2},
\end{eqnarray}
and then, Eq.~(\ref{h2}) is derived.

~\\

\end{appendix}

\end{widetext}


\begin{thebibliography}{0}
\bibitem{Silaev0} To justify the result from the Eilenberger equation, Ref.~\onlinecite{Silaev} also performed the diagrammatic formulation. Nevertheless, as referred to Sec.~IV~A in Ref.~\onlinecite{Silaev}, a special procedure that follows the exact step to derive the Eilenberger equation by first taking the commutation between the self-energy and $\tau_3$-Green function and applying the quasiclassical approximation afterwards, was applied to handle the calculation within the diagrammatic formalism. Consequently, the applied diagrammatic formalism approach with this procedure becomes exactly same as the Eilenberger equation and is not a separate nontrivial techinique.
\end{thebibliography}
\end{document}